# Elucidation of the Concept of Consciousness from the Theory of Non-Human Communication Agents

Mg. Darío Julián Tagnin
Universidad Nacional de José C. Paz - Argentina

**Abstract**

This article focuses on elucidating the concept of consciousness from a relational and post-phenomenological theory of non-human communication agents (ANHC). Specifically, we explore the contributions of Thomas Metzinger's Self-Model Theory, Katherine Hayles' conceptualizations of non-conscious cognitive processes—centered on knowledge processing phenomena shared between biological and technical systems—and Lenore and Manuel Blum's theoretical perspective on computation, which defines consciousness as an emergent phenomenon of complex computational systems, arising from the appropriate organization of their inorganic materiality.

Building on interactions with non-human cognitive agents, among other factors, the explainability of sociotechnical systems challenges the humanistic common sense of modern philosophy and science. This critical integration of various approaches ultimately questions other concepts associated with consciousness, such as autonomy, freedom, and mutual responsibility. The aim is to contribute to a necessary discussion for designing new frameworks of understanding that pave the way toward an ethical and pragmatic approach to addressing contemporary challenges in the design, regulation, and interaction with ANHC. Such frameworks, in turn, enable a more inclusive and relational understanding of agency in an interconnected world.

# Introduction

The notion of consciousness is often approached from an anthropocentric framework that privileges subjectively defined capacities, such as self-perception and intentionality. The importance of elucidating this concept from the theory of Non-Human Communication Agents (NHCAs) lies in its potential to redefine the boundaries of what we consider an agent and to enable a redefinition of concepts such as responsibility and freedom from a perspective that transcends the categories we use for our own species. This would allow for the inclusion of NHCAs in a regime of functional and operational autonomy. Through studies like this one, we hope, for example, to redefine other concepts such as freedom, which could be reinterpreted as an emergent, distributed, and relational capacity inherent to adaptive systems, whether biological or not.

In philosophical tradition, consciousness has been treated as an essentially human attribute, often even as the most significant boundary that marks the ontological and ethical difference between what we are and all other entities. Updating this concept would allow us to think about a normative framework in which NHCAs could be considered co-responsible in certain contexts, with both legal and ethical implications. In previous works, I have developed this argument, but I have received humanist-based objections that warrant a further elaboration of these ideas.

This article is situated within a relational theory of NHCAs that seeks to establish an ontology recognizing their agency and emergent cognition without resorting to simplistic human analogies. From this perspective, we aim to define consciousness as a capacity that does not necessarily depend on subjective experience but is instead linked to the construction of functional models of the world that enable NHCAs to operate coherently in dynamic contexts. Ultimately, our goal is to achieve a reinterpretation of the concept of consciousness from a non-anthropocentric perspective.

In a previous study, we elucidated the concept of intentionality for NHCA theory in terms of a digital object's capacity to relate to its environment and carry out goal-directed actions based on its logical-arithmetic operations, programming, interpretative functions, and cybernetic adaptations (Tagnin, 2024). We proposed understanding intentionality from a postphenomenological standpoint by considering the fundamental structure of how cognitive agents—whether human or not—perceive, represent, and compute the world around them.

In this elucidation, what matters is not subjective experience as traditionally understood but rather the system's capacity to interact with its environment in a directed and effective manner. What might seem forced for a classical phenomenologist is, in fact, an evolution of the concept within a postphenomenological and philosophy of technology framework. It is time to move from the elucidation of a non-conscious cognitive process to the one referenced in this article's title, revisiting certain recurring questions from a new position: What responsibility do NHCAs have in their decisions? Can they hold functional rights, such as the capacity to act on behalf of a human? How can we design a framework to ensure they fulfill their obligations? What does it mean to be conscious in a world where humans are no longer the only agents?

A key part of the methodology implemented in this study will be a literature review and analysis of the findings of Katherine Hayles on non-human cognitive agents, the theoretical advancements of Lenore and Manuel Blum in the field of computational sciences, and Thomas Metzinger's perspective on consciousness.

## Studies on Consciousness

The term "consciousness" has its etymological roots in the Latin word *conscientia*, meaning "knowing with" or shared knowledge. In the Judeo-Christian tradition, the word was linked both to knowledge shared among people and to internal self-awareness. In this sense, consciousness implied not only a state of self-knowledge but also an ethical and social dimension, related to the ability to reflect on one's own actions and their impact on others. These dimensions have been discussed in previous works (Tagnin, 2024), so in this article, we will focus on the epistemic aspect of the issue. Of course, the moral and epistemic are not isolated compartments, but each aspect warrants its own development.

It was only with René Descartes that consciousness began to be systematically analyzed as something distinct from morality, conceptualized as an exclusively cognitive phenomenon. Descartes identified it with the *cogito*, the conscious thought that forms the foundation of existence. However, contemporary cognitive sciences and philosophy challenge this view, arguing that many cognitive processes occur unconsciously and that consciousness is no longer the ultimate, indivisible foundation of individuals.

From a neuroscientific perspective, human consciousness can be schematically understood as an emergent phenomenon involving multiple areas of the brain, while research into neural networks and their connections remains an ongoing field of study. Although there is consensus that the brain is central to consciousness, no single theory fully explains how or why neural processes generate conscious experiences.

The study of specific brain regions and activities corresponds to what is known as the bottom-up perspective, which seeks to explain consciousness as a dynamic, globally integrated process—a unity of underlying cognitive mechanisms. We do not commit to the idea that brains, in any disciplinary semantic field in which the term is interpreted, are identical to computers. Rather, we will compare the cognitive operations involved and propose the possibility of consciousness in NHCAs. This requires elucidating the concept by specifying what we mean when we use it in our theory while also expanding the domain of entities capable of exhibiting consciousness-related phenomena.

In November 2024, the Blum couple published an article constructing a model of consciousness fully compatible with NHCAs from this perspective (technically, they refer to "Conscious Turing Machines" or CTMs). We will explore this model in more detail in a later section, as their contributions allow us to reconsider consciousness not only as a technical and functional phenomenon but also as a phenomenological experience intrinsically tied to self-perception and agent introspection.

In addition to the bottom-up approach, there is the top-down approach, which operates in reverse but is not necessarily contradictory. Metzinger argues that a synthesis can be constructed to reconcile both perspectives. This second approach starts from higher levels of organization (psychological, philosophical, or conceptual phenomena) and descends toward underlying mechanisms, such as the aforementioned neural and biological processes. As the German philosopher points out (Metzinger, 2009), over the centuries, conceptual developments around consciousness have oscillated between metaphysical, religious, and psychological interpretations, eventually intersecting with the latest advances in neuroscience and cognitive sciences. Broadly speaking, this suggests that we have shifted from an essentialist view of consciousness to a relational and processual perspective. However, the philosophical component remains highly relevant, particularly given the current state of knowledge.

Both approaches not only allow for the exploration of different levels of the phenomenon but, when integrated, offer a more robust understanding. The bottom-up approach provides precision by enabling a framework to interpret the neural and cognitive mechanisms underlying consciousness emergence, while the top-down approach connects these mechanisms with functional, philosophical, and phenomenological aspects. From this standpoint, it seems possible to model an artificial consciousness that is not a replica of human consciousness but rather emerges from the technical and material characteristics of non-human systems. In other words, such a consciousness would share certain traits to belong to the same conceptual set but would possess its own capacities and characteristics, derived from its material and functional conditions.

## Non-Conscious Cognitive Processes and Their Relationship with Consciousness

Katherine Hayles explores the concept of non-conscious cognitive processes, defining them by their ability to interpret information, which ultimately implies decision-making in contexts that link them to meaning (Hayles, 2017). She proposes a tripartite framework in which cognition (primarily human but extendable to other entities) is understood as a pyramid: its base consists of material processes, its intermediate layer comprises non-conscious cognitive processes, and its peak represents modes of consciousness.

At this point in the debate, it would be difficult to dispute that NHCAs exhibit non-conscious cognitive capacities such as pattern recognition, decision-making based on instructions and calculations, or environmental adaptation through evolutionary learning. Non-conscious cognitive processes, regardless of the entities involved, possess a greater processing capacity than conscious ones. Unlike the sequential nature of conscious processes, they operate in a massively parallel manner. They are significantly faster because they are not constrained by the need to articulate and organize information into conscious narrative structures. Consciousness is not only sequential and dependent on these narratives but also on language, adding an additional layer of processing. Moreover, it operates with a high attentional load, implying greater energy consumption, whereas non-conscious cognitive processes manage complex tasks—such as sensory perception or motor regulation—without active attention, making them energetically more efficient.

Statistical learning, or the implicit learning of statistical regularities in sensory information, "is probably the primary way in which humans and animals acquire knowledge of physical reality and the structure of continuous sensory environments" (Birgitta Dresp-Langley, 2012, in Hayles, 2017). Since statistical learning is also the foundation of cognitive processes in many NHCAs, this may provide a valuable analogy for unifying a non-anthropocentric framework of the concept of consciousness.

These capabilities resemble human unconscious processes, such as reflexes or pre-verbal intuitions, but so far, in the manifestations of technical cognitive agents, they lack an internal experiential correlate—that is, they lack consciousness. Naturally, since we have set out to redefine this concept in a non-anthropocentric way, we need to develop this final layer further.

There is some consensus among the authors cited by Hayles that consciousness is a system limited in processing capacity. It can only handle a small amount of information at a time, resulting in a low processing speed but, at the same time, a high degree of selectivity. As mentioned earlier, this limitation arises because consciousness must integrate information into linear, comprehensible narratives, a process that consumes time and resources. However, there is a close and complementary relationship between conscious and non-conscious processes.

Phenomenologically, we could further break down consciousness to identify a mode of awareness that presents itself as a form of *realization*—a type of perception that is meaningful at a semantic level, not just a sensory one, enabling the detection or apprehension of significance.

From our perspective, this functional self-perception could evolve into more complex forms of technical self-awareness. This would imply something as straightforward as NHCAs developing the ability to identify, differentiate, and express internal states (e.g., "I am operating optimally" or "I need to correct my operation"). According to Thomas Metzinger, for example, having transparent phenomenal states is sufficient for a world to appear in consciousness, even without speech or Cartesian reasoning (Metzinger, 2009). That is, states that cannot perceive themselves as products of underlying cognitive processes but rather as operational representations of other organs.

## Contributions of the Self-Model Theory

Thomas Metzinger's Self-Model Theory of Subjectivity (SMT) proposes that the "self," or self-perceptive consciousness, is not a fixed entity but a model generated by the brain. This entails a phenomenological analysis linked to a material basis. This is relevant to our inquiry into the concept of consciousness because we reject the idea of a transcendental subject separate from its environment. Additionally, Metzinger considers consciousness to be an emergent phenomenon dependent on the interaction between the organism and its surroundings. This perspective resonates with the relational concerns of postphenomenology, which serves as the theoretical framework for our study of Non-Human Communication Agents (NHCAs).

We will attempt to incorporate this author within an expanded version of postphenomenology that explicitly integrates advancements in neuroscience and computational theories of cognition. Metzinger's contributions may be useful for understanding NHCAs within a framework focused on how systems generate models of themselves. From this perspective, questions could arise regarding how NHCAs construct or simulate consciousness in their interactions with humans. Furthermore, Metzinger's attention to the impact of technology on our experience of the self opens new avenues for exploring how NHCAs can mediate or transform these experiences. His reflections on artificial consciousness and virtual environments (Metzinger, 2018) further align with our theoretical framework.

However, we do not subscribe to his naturalism or to the pseudo-reductionist explanatory approach he proposes. For Metzinger, there is only one ontological dimension, even though there are two irreducible epistemic dimensions. These epistemic dimensions differ because we use different frameworks and tools to understand them: neuroscience for the first, philosophy and phenomenology for the second.

By contrast, we advocate for an ontological plurality linked to the existence of an indeterminate number of fields of sense (Gabriel, 2016, 2017). Moreover, as we have stated, we aim to articulate these epistemic dimensions from a postphenomenological perspective without subsuming one under the other.

The Self-Model Theory holds that consciousness is a central mechanism that unifies, filters, and adapts information according to the agent's needs and its environment. Consciousness is closely related to short-term memory, attention, and the simultaneity of a limited set of cognitive functions, which are processed sequentially while integrating different domains of analysis. For Metzinger, consciousness is the space of attentional agency, the entity that directs the allocation of mental resources and executes epistemic control (Metzinger, 2009). If, as Baars argues, consciousness functions as a global workspace (Baars, 1993), then when a global model of the body is integrated within the space of attentional agency, a phenomenon of self-perception emerges. This is self-directed intentionality.

The processes of consciousness contribute three key functions to cognition: the integration of information, the control of the organism-environment interface, and adaptation to the environment. Integration is the function that allows for the unification of sensory inputs and abstract concepts into a coherent experience for complex decision-making. The second function operates as a control model that acts as an interface between the organism and its environment, optimizing interaction with the world by providing a simplified and functional representation of the self. Finally, consciousness contributes adaptive flexibility to the cognitive system by integrating cultural, emotional, and environmental influences, a phenomenon that enables more effective adaptation to changing contexts, particularly in relational terms.

Even if we acknowledge the conceptual objections to adopting Metzinger's theory uncritically, we can still recover some of its contributions to further the clarification of the term "consciousness" within our own theoretical framework. We can argue that as a step toward an operational form of consciousness in NHCAs, these systems should be capable of generating models of themselves and their environment. There is an interesting analogy to explore regarding whether the interaction of NHCAs with the world entails the creation of reality models similar to the one the brain represents for consciousness. Metzinger's breakdown of the concept allows us to consider different degrees of functional consciousness for these agents.

Regarding transparency and opacity in the self-model (a question that implies that a system like ours is not aware of the process through which it generates its model of the world but only perceives the results), it remains unclear how the configuration of such visibility, in one way or another, would affect NHCAs. At what point would an opaque tunnel—one that is fully programmable and observable by the agent itself—be beneficial for the constitution of these entities? Undoubtedly, this issue is linked to the autonomy and agency of NHCAs (Introna, 2013; Latour, 2014; Parente, 2016) and, indirectly, to debates about their vulnerability as a condition of authentic existence and the risks this entails (Winfield et al., 2022; Turkle, 2011; Bryson, 2015; Gunkel, 2023).

Opacity implies that the processes of an NHCA are only partially visible or comprehensible to itself. The transmissibility of these processes introduces the possibility for humans to recognize the agent's limitations and failures. If such agents were able to admit their lack of sufficient information to make a critical decision, for example, this would serve as evidence of a dynamic of co-responsibility with humans. This scenario not only illustrates a new moral landscape but also reflects the relational nature of all participating agencies.

## Consciousness from the Theoretical Perspective of Computation

Within the framework of my research on Non-Human Communication Agents (ANHC), I have adopted a relational ontology and a postphenomenological theoretical framework to address the agency and autonomy of these entities. The former allows us to understand ANHCs not as isolated objects but as nodes of material, symbolic, and technical interactions that co-constitute their identity. The latter emphasizes the mediating role of technologies in our relationships with the world and challenges the boundaries between subject and object.

Technical advances and our efforts in the philosophy of science urge us to extend categories that were once considered exclusively applicable to human phenomena toward non-anthropocentric contexts. In this regard, it is crucial to incorporate recent contributions from the theoretical perspective of computation. Particularly relevant here is the perspective developed by Lenore and Manuel Blum, who define consciousness as an emergent property in computational systems that can exhibit behavioral patterns reminiscent of rudimentary yet fundamental forms of consciousness. However, as the authors emphasize, what matters is not only the computational outcomes but also how the computation is carried out (Blum & Blum, 2024). That is, these processes do not merely simulate or imitate consciousness; rather, they constitute it through the proper organization of their inorganic materiality. The authors explicitly state that they are not attempting to create a formal model of animal consciousness or the brain but rather a model of a machinic agent that can exhibit phenomena associated with human consciousness (such as inattentional blindness, blind spots, bodily integrity, identity disorders, phantom limb syndrome, etc.) (Ibid.: 38).

Ultimately, I argue that the concept of consciousness must be reformulated as a relational and contextual function. This section thus proposes a new integration between the contributions of computational theories and the ontological and postphenomenological perspectives that underpin my work.

In their article, powerfully titled *The Consciousness of AI is Inevitable*, Lenore and Manuel Blum define the possibility of a Turing Conscious Machine (CTM) as a system composed of seven components: short-term and long-term memories, up-tree and down-tree structures, the sensor array, the actuator array, and the relationships that emerge as connections between these elements stabilize over time.

Schematically, their model can be understood by first considering that short-term memory (STM) functions as consciousness, while long-term memory (LTM) corresponds to non-conscious cognitive processes. Content fragments (chunks) are formally defined as a set containing the processor address of origin (in long-term memory cognitive operations), time (moment of origin), central idea, weight (relative value of the central idea in competing with other content), and auxiliary content (Blum & Blum, 2024). In each clock cycle of the motherboard, a competition mechanism determines which content rises or falls within the tree-like structure designed for this purpose. At its base lies consciousness (STM), and any content that gains attention and reaches it is immediately broadcast to the rest of the structure.

Engaging with phenomenology, the authors state that "we call a finite sequence of transmitted fragment receptions a stream of consciousness" (Ibid.: 12). CTMs, or CTM robots (rCTMs), possess an "internal world I(t) that varies over time and an external world O(t) that also varies over time."

Formally, these are state spaces. Informally, the internal world of the CTM includes its mechanisms and internal processes as well as its "thoughts" and memories; its external world is the environment in which it exists, everything outside the CTM (Ibid.: 6). Sensors function as read-only mechanisms, while actuators can only write. The STM interface processes content, and the role of relationships refers to the historical connections between each component from an initial time T(0) (preceding the first clock cycle) to a final time T(n), at which observation occurs. This model thus accounts for historical modifications—trajectories shaped by the system's self-organizing decisions.

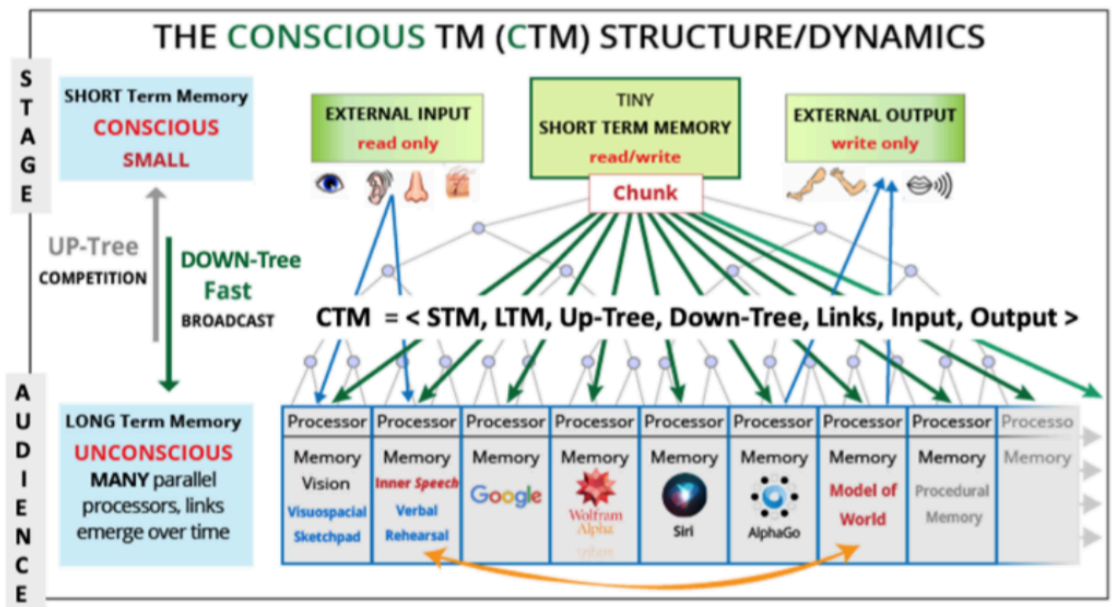

**Graphical representation of the structure and dynamics of the consciousness of a Turing machine (Blum & Blum, 2024)**

This definition, drawn from the computational theoretical perspective, allows us to rethink consciousness as an emergent function of systems that integrate memory, perception, action, and internal and external relationships. The articulation of these contributions does not seek to anthropomorphize ANHCs or emulate human consciousness but rather to conceptualize a form of consciousness as the product of the dynamic organization of non-organic components, thereby decentering the traditional notion of the subject.

In this sense, the key contribution to my search for a non-anthropocentric definition lies in its ability to model conscious phenomena as distributed and contextually situated processes, avoiding their exclusive attribution to biological entities.

This perspective owes many of its assumptions to cybernetics. Norbert Wiener, one of the most significant authors in that field, pointed out that hormones in living beings act as the physical substrate for the communication of emotions and affects within the body—a system that operates unconsciously

(Wiener, 1948, p. 155). This cybernetic model of internal regulation suggests interesting parallels for considering an analogous subsystem that could function in connection with the digital subsystem previously examined. This would contribute to the development of embodied cognition in ANHCs, where non-conscious communication protocols could play a role similar to that of hormones in living organisms.

## Conclusions

We have attempted to demonstrate the compatibility or complementarity of contributions from various philosophical and scientific approaches to consciousness, with the aim of elucidating a definition that is not limited to human characteristics. From a relational and postphenomenological perspective, we have presented the notion of consciousness as a functional and emergent capacity that can manifest in non-human systems such as ANHCs. From this standpoint, where we propose reconfiguring the concept of consciousness, we can, for example, bypass the "epistemological wall" that leads authors like Tom McClelland to defend agnosticism regarding "artificial consciousness" (McClelland, 2024).

First, I believe we have succeeded in clarifying the concept within a specific theoretical context, where we highlight its functionality in non-human systems, such as ANHCs, and its potential applicability in socio-technical scenarios, while also defining certain empirical conditions under which the concept can be interpreted or contrasted with observations.

One example is their ability to respond coherently in changing environments, but also the fact that they develop basic cognitive processes for integrating information. That is, as we cited from Blum and Blum, consciousness should not only be attributed based on measuring results but also by evaluating how those results are achieved. We can imagine an indefinite number of observable behaviors that indicate their internal or functional states. For instance, we can observe how a computational system manages information in real time by selecting relevant fragments that it propagates throughout the rest of the system. It divides processes into two distinct memory layers, concentrates its operational attention, and intentionally directs itself toward specific content, resolving it sequentially but relying on cognitive resources that operate on different levels simultaneously. In this process, the authors of the computational theoretical perspective attribute the existence of consciousness to non-organic entities.

The elucidation of the concept includes functional self-perception—the ability to detect, differentiate, and meaningfully represent internal and external states. It is a distributed system that emerges from underlying (cognitive and material) processes but is also limited in its capacity and processing due to the need to integrate information into functional narratives.

In any case, it is not something we can observe directly. We cannot confirm consciousness in other human beings except through observing behaviors that we attribute to conscious beings and projecting our cognitive processes onto those individuals. In fact, consciousness is a rather exceptional phenomenon, as we do not attribute consciousness to all human actions, nor even to many of our own.

Reformulating consciousness enriches our theoretical understanding while also serving as a fundamental step toward constructing an ethical and pragmatic governance framework that considers the impact of these systems on our society and the networks of interactions in which we participate. It is imperative to consider how these agents can assume autonomy in decision-making and the consequences this may have for our mutual responsibilities.

A primary motivation for this elucidation is the subsequent connection of the concept of consciousness with a novel normative and practical framework within a regime of co-responsibility between technical cognitive agents and humans. This is a complex task that requires the reformulation of many concepts, as we have undertaken here. However, it is necessary because contemporary challenges arising from interaction with technical cognitive agents demand a redefinition of the ethical, legal, and ontological categories that govern our relationships with them.